\documentclass{andp2012}
\usepackage[english]{babel}

\usepackage{kantlipsum,cuted}

\usepackage[normalem]{ulem}
\usepackage{amsmath,amssymb}
\usepackage{multirow}
\usepackage{xcolor,soul}
\usepackage{srcltx}
\usepackage{hyperref,graphicx}

\title{Anderson localization in dissipative lattices}
\author[S: Longhi]{Stefano Longhi \inst{1,}\footnote{Corresponding author\quad E-mail:~\textsf{stefano.longhi@polimi.it}}}
\address[1]{Dipartimento di Fisica, Politecnico di Milano,Piazza L. da Vinci 32, I-20133 Milano, Italy and IFISC (UIB-CSIC), Instituto de Fisica Interdisciplinar y Sistemas Complejos, E-07122 Palma de Mallorca, Spain}
\shortauthors{S. Longhi }
\begin{abstract}
\small
Anderson localization predicts that wave spreading in disordered lattices can come to a complete halt, providing a universal mechanism for {dynamical localization}. In the {\color{black} one-dimensional Hermitian Anderson model with uncorrelated diagonal disorder}, there is a one-to-one correspondence between dynamical localization and spectral localization, i.e. the exponential localization of all the Hamiltonian eigenfunctions.
This correspondence can be broken when dealing with disordered dissipative lattices.
Recently, it has been shown that when the system exchanges particles with the surrounding environment and random fluctuations of the dissipation are introduced, spectral localization is observed but without dynamical localization. Such previous studies considered lattices with mixed conservative (Hamiltonian) and dissipative dynamics, and were restricted to a semiclassical analysis. However, Anderson localization in purely dissipative lattices, displaying an entirely Lindbladian dynamics, remains largely unexplored. Here we consider the purely-dissipative Anderson model in the framework of a Lindblad master equation and show that, akin to the semiclassical models with conservative hopping   and random dissipation, one observes dynamical delocalization in spite of strong spectral localization of all eigenstates of the Liouvillian superoperator. This result is very distinct than delocalization observed in the Anderson model with dephasing effects, where dynamical delocalization arises from the delocalization of the stationary state of the Liouvillian superoperator.
\end{abstract}
\shortabstract
\begin{document}
\maketitle

\section{Introduction}

Anderson localization \cite{r1} describes the absence of wave spreading (diffusion) in lattices with on-site uncorrelated potential disorder, that is observed above a critical disorder strength.
Such a universal phenomenon occurs both in quantum and classical systems and is of wide impact in various fields of physics with major implications in quantum and condensed-matter physics, Bose-Einstein condensates, photonic and quantum technologies  \cite{r2,r3,r4,r5,r6,r7,r8,r9,r10,r11,r12,r12b,r13,r14,r15,r16,r17,r17b,r18,r19,r20,r21,r22,r23,r24,r25,r26}.
In his seminal paper \cite{r1}, Anderson analyzed the problem of propagation of a single quantum
particle in a disordered potential under unitary time evolution, i.e. in the coherent Hamiltonian limit. In this regime the localization phenomenon arises from an intricate interference effect, where the destructive interference of many amplitudes leads to the exponential localization of the wave functions. Nowadays the single-particle dynamics of the Anderson model is well understood and several rigorous mathematical results have been established (see e.g. \cite{r27}
for a review).  In particular, in one-dimensional lattices suppression of wave spreading, i.e. {\it dynamical localization}, occurs at any infinitesimally small strength of uncorrelated disorder and it is intimately related to the exponential localization of the Hamiltonian eigenfunctions, i.e. to {\it exponential spectral localization} \cite{r27}. 
Indeed, in conservative systems under weak conditions \cite{r28}, and surely for the Anderson model \cite{r29}, exponential spectral localization is a necessary and sufficient condition to suppress wave spreading in the lattice, i.e. an equivalence can be established between spectral and dynamical localization. \par
Such a simple scenario of Anderson localization is deeply modified when the system interacts with its environment
 exchanging energy and/or particles \cite{r30,r31,r32,r33,r34,r35,r36,r37,r38,r39,r40,r40b}. Intuitively, interaction with an environment results in decoherence, which 
washes out interference effects and thus destroys Anderson localization. Indeed, several works have shown that inclusion of dephasing  or continuous measurements in the Anderson model
 gives way to delocalization \cite{r30,r31,r32,r34,r40,r40b}, while special engineering of system-environment coupling could drive the system toward a localized state with tunable localization \cite{r36,r37,r38}.
 In open systems, the dynamics can be described using a master
equation in Lindblad form for the reduced density operator or an effective non-Hermitian Hamiltonian, which neglects quantum jump terms in the master equation (see for instance \cite{r40c,r41,r42}). The non-Hermitian approach
can be regarded as a semiclassical limit of the full quantum dynamics \cite{r41,r42}; {\color{black} in a quantum-jump approach to dissipative dynamics, it also describes quantum evolution of open systems under continuous measurements and post-selection \cite{RR1,RR2}{}}. Recently, there has been a renewed interest on the Anderson localization problem in dissipative systems using an effective non-Hermitian approach \cite{r43,r44,r45,r46,r47,r48,r49}. In particular it has been theoretically predicted \cite{r43,r47,r48,r49} and experimentally demonstrated \cite{r48} that random fluctuations solely in the dissipation, introduced by uncorrelated on-site imaginary potential disorder in the effective non-Hermitian Hamiltonian, could exponentially localize all eigenstates (spectral localization), similar to the original case without dissipation that Anderson considered. However, rather surprising wave spreading is not suppressed in spite of spectral localization, and the evolution of normalized occupation probabilities in the lattice  is characterized by stochastic quantized jumps among exponentially-localized eigenstates located at distant sites and displaying increasing lifetimes \cite{r47,r48,r49}, resulting in a diffusive-like spreading.\par
In all previous studies of Anderson localization the dynamics is mixed,  with characteristics of both Hermitian Hamiltonian (conservative) evolution and dissipative Lindbladian evolution. 
However, Anderson localization  in lattices with purely dissipative couplings \cite{r49b,r50,r52b}, thus displaying a purely  dissipative Lindbladian dynamics, remains largely unexplored. In this work we introduce a purely-dissipative Anderson model, where the entire dynamics is dissipative Lindbladian, and unravel that, similar to  the semiclassical dissipative models displaying mixed dynamics \cite{r47,r48,r49}, dynamical delocalization is observed in spite of  the  exponential localization of all eigenstates of the Liouvillian superoperator. We also point out that the physical origin of delocalization observed in the purely-dissipative Anderson model arises from the stochastic quantized jumps of localized states at distant sites in the lattice, displaying different lifetimes \cite{r48,r49}, and it is thus very distinct than delocalization found  in the dephasing Anderson model \cite{r39}, which originates from the delocalization of the stationary state of Liouvillian superoperator.

\section{Dissipative Anderson models}
Let us start by introducing a rather general class of dissipative Anderson models that will guide us through the
general discussion and illustrate our results. For the sake of definiteness, let 
us consider a one-dimensional chain of $(2L+1)$ optical cavities or waveguides, described by the bosonic optical modes $\hat{{a}}_l$, which can be coupled either conservatively
at a rate $J$ or dissipatively at a rate $\Gamma$ \cite{r49b,r50,r52b,r51,r52} (Fig.1). {\color{black} In all analytical and numerical analysis, we consider a finite linear chain with open boundary conditions.} Dissipative coupling indirectly couples the cavities in the lattice
through an intermediate common reservoir, and can be experimentally implemented in several photonic platforms \cite{r50,r52b}. We indicate by $\delta \omega_l$ the 
resonance frequency detuning of the $l$-th optical cavity mode from a reference frequency $\omega_0$, and by $\gamma_l$ the cavity mode loss rate.  
Stochasticity in the system can be introduced by considering either uncorrelated disorder for the resonance frequencies $\delta \omega_l$ or cavity losses $\gamma_l$. We also include
in the dynamics pure dephasing effects \cite{r40,r53}, which deprive the system of coherence but leave the energy (photon number) conserved. Typically, we will consider the thermodynamic limit $L \rightarrow \infty$, such that the notion of spectral localization used in the theory of Anderson localization becomes meaningful and edge effects in the spreading dynamics can be neglected.\\
The dynamics of the lattice can be described by the general Lindblad master equation for the system density operator $\hat{\rho}$, which reads  \cite{r36,r41,r42,r50,r51,r52}
\begin{eqnarray}
\frac{d}{dt} \hat{\rho} & = &  \mathcal{L} \hat{\rho}=-i [ \hat{H}, \hat{\rho} ]+ \\
& + & \sum_l  \left( \Gamma \mathcal{D} [ \hat{z}_l] \hat{\rho}+\gamma_l  \mathcal{D} [ \hat{a}_l] \hat{\rho} 
+\gamma_{ph}  \mathcal{D} [ \hat{a}^{\dag}_l \hat{a}_l] \hat{\rho} \right). \nonumber
\end{eqnarray}
The first term on the right hand side of Eq.(1)  describes the Hermitian (unitary) Hamiltonian dynamics due to
conservative couplings between adjacent cavities at a rate $J$ and governed by the Hamiltonian
\begin{equation}
\hat{H}= - J \sum_l  \left( \hat{a}^{\dag}_{l+1} \hat{a}_l+ \hat{a}^{\dag}_{l} \hat{a}_{l+1} \right)+ \sum_l \delta \omega_l \hat{a}^{\dag}_{l} \hat{a}_l
\end{equation}
whereas the last terms on the right hand side of Eq.(1) describe the dissipative part of the Liouvillian superoperator $\mathcal{L}$. The dissipative dynamics comprises three terms:\\
(i)  the dissipators 
\begin{equation}
\mathcal{D}[\hat{z}_l]\hat{\rho}=\hat{z}_l \hat{\rho} \hat{z}_{l}^{\dag}-\frac{1}{2} \left(  \hat{z}_l^{\dag} \hat{z}_l \hat{\rho} + \hat{\rho} \hat{z}_l^{\dag} \hat{z}_l \right)
\end{equation} 
resulting from the nonlocal jump operators  $z_l=\hat{a}_l+\hat{a}_{l+1}$ and describing dissipative coupling between adjacent cavities at a rate $\Gamma$;\\
(ii)  the dissipators 
\begin{equation}
\mathcal{D}[\hat{a}_l] \rho=\hat{a}_l \hat{\rho} \hat{a}_{l}^{\dag}-\frac{1}{2} \left( \hat{a}_l^{\dag} \hat{a}_l \hat{\rho} + \hat{\rho} \hat{a}_l^{\dag} \hat{a}_l \right) 
\end{equation}
 resulting from the jump operators $\hat{a}_l$ and describing cavity losses at rates $\gamma_l$; \\
 (iii) and the dissipators  
 \begin{equation}
 \mathcal{D}[\hat{a}^{\dag}_l \hat{a}_l] \rho= \hat{a}^{\dag}_l \hat{a}_l \hat{\rho} \hat{a}_{l}^{\dag}\hat{a}_l -\frac{1}{2} \left(  \hat{a}_l^{\dag} \hat{a}_l  \hat{a}^{\dag}_l \hat{a}_l \hat{\rho} + \hat{\rho} \hat{a}_l^{\dag} \hat{a}_l \hat{a}_l^{\dag} \hat{a}_l \right)
 \end{equation}
 resulting from the jump operators $\hat{a}^{\dag}_l \hat{a}_l$ and describing local dephasing decay at a rate $\gamma_{ph}$.\\
 The Lindblad master equation can be viewed as the time evolution of the system which is continuously monitored by an environment \cite{r41}. This interpretation is at best captured by rewriting Eq.(1) 
 in the equivalent form
 \begin{eqnarray}
 \frac{d}{dt} \hat{\rho} & = & -i \left(  \hat{H}_{eff} \hat{\rho} - \hat{\rho}  \hat{H}_{eff}^{\dag} \right)+ \Gamma  \sum_l  (\hat{a}_l+\hat{a}_{l+1}) \hat{\rho} (\hat{a}_l^{\dag}+\hat{a}_{l+1}^{\dag}) \nonumber \\
 & + & \sum_l \gamma_l \hat{a}_l \hat{\rho} \hat{a}_l^{\dag}+\gamma_{ph} \sum_l \hat{a}_l \hat{a}_l^{\dag} \hat{\rho} \hat{a}_l \hat{a}_l^{\dag}
 \end{eqnarray}
 where
 \begin{eqnarray}
 \hat{H}_{eff} & = & \hat{H}-\frac{i}{2} \Gamma \sum_l (\hat{a}_l^{\dag}+\hat{a}_{l+1}^{\dag}) (\hat{a}_l+\hat{a}_{l+1}) \nonumber \\
 & - &  \frac{i}{2} \sum_l \gamma_l \hat{a}_{l}^{\dag} \hat{a}_l-\frac{i}{2} \gamma_{ph} \sum_l \hat{a}_l^{\dag} \hat{a}_l \hat{a}_l^{\dag} \hat{a}_l
 \end{eqnarray}
 in the effective non-Hermitian Hamiltonian.
  \begin{figure}
  \includegraphics[width=\linewidth]{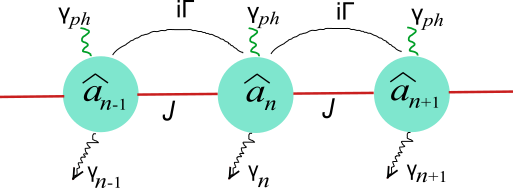}
  \caption{Schematic of a one-dimensional tight-binding lattice of coupled optical cavities. $J$ and $\Gamma$ are the conservative and dissipative coupling rates between adjacent cavities, respectively, $\gamma_n$ is the decay rate of the $n$-th  lossy cavity, and $\gamma_{ph}$ is the dephasing rate. The cavity resonance frequencies deviate from the reference value $\omega_0$ by $\delta \omega_n$. The purely-dissipative Anderson model considered in this paper, displaying purely Liouvillian dynamics, corresponds to $J=\delta \omega_n=\gamma_{ph}=0$; the dephasing Anderson model, displaying mixed Hamiltonian and Liouvillian dynamics, corresponds to $\gamma_n=\Gamma=0$; finally the Anderson model with stochastic dissipation, displaying mixed Hamiltonian and Liouvillian dynamics, corresponds to $\Gamma=\delta \omega_n=\gamma_{ph}=0$.}
\end{figure}
  The temporal evolution of density operator, described by Eq.(6), is basically composed by two terms. The first term on the right hand side of Eq.(6), involving the non-Hermitian Hamiltonian $\hat{H}_{eff}$, describes a non-unitary dynamics that accounts for the conservative (Hermitian) dynamics corrected by the continuous losses of energy,
information, and coherence of the system into the environment. The other terms on the right hand side of Eq.(6) are dubbed the quantum jump terms as they describe the effect of the
measurement on the state of the system:  in a quantum trajectory
approach  such terms are responsible for the abrupt stochastic
change of the wave function \cite{r41}. The evolution of the density operator neglecting the quantum jump terms provides a semiclassical limit of the underlying dynamics \cite{r41}; {\color{black} it also correctly captures the full quantum evolution of the system under continuous measurements and post-selection \cite{r41,RR1,RR2}.}\\
The master equation (6), along with its semiclassical limit obtained by neglecting quantum jumps and working with the effective non-Hermitian Hamiltonian $\hat{H}_{eff}$, describes a variety of dissipative Anderson models, some of which have been studied in previous works. Clearly, when the lattice is isolated, i.e. for $\Gamma=\gamma_l=\gamma_{ph}=0$, one obtains the usual (Hermitian) Anderson model, originally studied by Anderson \cite{r1}, which displays both dynamical localization and exponential spectral localization of the eigenstates for any strength of uncorrelated disorder of the resonances $\delta \omega_l$. The exponential localization of the wave functions involves complicated interference effects of waves, which requires coherence in the system. When considering dissipative terms, Eq.(6) can describe three interesting situations:\\
\\
1) The {\it dephasing Anderson model.} This model extends the ordinary Anderson model by considering dephasing effects, which introduce decoherence in the systems but do not change the particle number. This model, which is obtained from Eq.(6)  by letting $\Gamma=\gamma_l=0$, has been studied in previous works (see e.g. \cite{r38,r40,r40b}). As expected, decoherence fully destroys the delicate interference phenomenon at the heart of Anderson localization: dephasing restores dynamical delocalization and the stationary state of the density operator corresponds to a maximally mixed state with delocalized excitation in the lattice. {\color{black} For this model, the Lindbladian and effective non-Hermitian descriptions yield markedly distinct dynamical behaviors, as shown in the Appendix A, owing to the fact that the jump operators describing dephasing effects are quadratic in the bosonic operators.}\\
\\
2) The {\it Anderson model with stochastic dissipation}. This model was introduced in recent works \cite{r43,r44,r45,r46,r47,r48,r49} in the framework of an effective non-Hermitian (semiclassical) approach and is obtained by letting $\delta \omega_l=\Gamma=\gamma_{ph}=0$ in Eq.(6). In this model a clean conservative (Hamiltonian) dynamics is considered, with hopping rate $J$ between adjacent sites, while disorder is introduced by uncorrelated random dissipation $\gamma_l$ at various lattice sites; mixed disorder in both $\gamma_l$ and $\delta \omega_l$ has been also considered. As shown in such previous works, exponential spectral localization of all eigenstates of $\hat{H}_{eff}$ is found, like in the original Anderson model. However, quite surprisingly the system exhibits counterintuitive propagation by quantized jumps between exponentially-localized states located around
distant sites \cite{r47,r48,r49}, allowing for spatial spreading of excitation along the lattice (dynamical delocalization) in spite of spectral localization. Such a hopping dynamics between distant localized states 
is rooted in the  different lifetimes of the exponentially-localized eigenstates of the non-Hermitian Hamiltonian, so as distant eigenstates with longer lifetimes can dominate the dynamics at long times \cite{r48}. We remark that the wording "quantized jumps", introduced in Ref.\cite{r49} to refer to the stochastic sequence of dominant eigenstates observed in the dynamical evolution of the non-Hermitian Hamiltonian, should not be confused with the quantum jumps in the quantum trajectory approach of the Lindblad master equation.\\
\\ 
3) The {\it purely-dissipative Anderson model}. The previous models 1) and 2) of Anderson localization in open systems, as well as other models studied in previous works and involving different forms of jump operators \cite{r36,r37,r38}, display a {\it mixed} dynamics, with characteristics of both Hermitian Hamiltonian evolution and dissipative Lindbladian evolution. Here we consider a different situation, the {\it purely-dissipative} Anderson model, which has been overlooked so far. In the purely dissipative Anderson model there is not any Hermitian Hamiltonian evolution, i.e. $\hat{H}=0$, and the interaction with the environment is dissipative without dephasing effects ($\gamma_{ph}=0$). The purely-dissipative Anderson model thus displays a purely dissipative Lindbladian  evolution \cite{r50,r52b}, and it is obtained from the general model [Eqs.(6) and (7)] by letting $J=\delta \omega_l=\gamma_{ph}=0$.\\
\\ 
In the following we will mainly focus our attention to the purely-dissipative Anderson model, unraveling its spectral and dynamical properties. {\color{black} In particular, we will show that for this model the Lindbladian and effective non-Hermitian descriptions 
basically yield the same dynamical behavior, owing to the fact that the jump operators are linear in the bosonic destruction operators}.  

\section{Purely-dissipative Anderson model: single-particle subspace dynamics}
The master equation of the purely-dissipative Anderson model reads
 \begin{eqnarray}
 \frac{d}{dt} \hat{\rho}  & =  & -i \left(  \hat{H}_{eff} \hat{\rho} - \hat{\rho}  \hat{H}_{eff}^{\dag} \right) \\
 &+ & \Gamma  \sum_l  (\hat{a}_l+\hat{a}_{l+1}) \hat{\rho} (\hat{a}_l^{\dag}+\hat{a}_{l+1}^{\dag}) 
 +  \sum_l \gamma_l \hat{a}_l \hat{\rho} \hat{a}_l^{\dag} \nonumber
 \end{eqnarray}
 where the effective non-Hermitian Hamiltonian
 \begin{equation}
 \hat{H}_{eff}=-\frac{i}{2} \Gamma \sum_l (\hat{a}_l^{\dag}+\hat{a}_{l+1}^{\dag}) (\hat{a}_l+\hat{a}_{l+1})-\frac{i}{2} \sum_l \gamma_l \hat{a}_{l}^{\dag} \hat{a}_l
 \end{equation}
 is entirely anti-Hermitian. The initial state is rather generally described by a mixture of pure states $| \psi_{\sigma} \rangle$ with probabilities $p_{\sigma}$, i.e.
 \begin{equation}
 \hat{\rho}(0)=\sum_{\sigma} p_{\sigma} | \psi_{\sigma} \rangle \langle \psi_{\sigma} |
 \end{equation}
with $\sum_{\sigma} p_{\sigma}=1$. 
 Owing to the nature of the jump operators, which are linear functions of the destruction operators $\hat{a}_l$, the dynamics of the system for an initial state involving a number of bosonic particles (photons) $\leq N$ is entirely enclosed in the sub-space of Hilbert space spanned by the Fock states 
 \[
 \frac{1}{ \sqrt{\prod_l n_l! }} \;  ... \; \hat{a}_{l-1}^{\dag n_{l-1}} \hat{a}_{l}^{\dag n_l} \hat{a}_{l+1}^{\dag n_{l+1}} \; ... \; | vac \rangle, 
 \] 
 where $n_l \geq 0$, $\sum_l n_l \leq N$ and $| vac \rangle$ is the vacuum state. To unravel the localization features of the dissipative Anderson model, let us consider first the dynamics of the density operator in the single-particle subspace ($N \leq1$); the most general case will be discussed in the next section. The single-particle subspace is described by $(2L+2)$ Fock states, corresponding to the localization of the photon in either one of the $2L+1$ cavities of the lattice and to the vacuum state $| vac \rangle$. We indicate the single-particle Fock states as $| l \rangle \equiv  \hat{a}_l^{\dag} |vac \rangle$ ($l=-L,-L+1,..,0,1,...,L$). 
The evolution equations for the density matrix elements $\rho_{n,m} \equiv \langle n | \hat{\rho} | m \rangle=\rho_{m,n}^*$, $\rho_{n,vac} \equiv \langle n | \hat{\rho} | vac \rangle =\rho_{vac,n}^*$ and $\rho_{vac,vac} \equiv \langle vac | \hat{\rho} | vac \rangle$ can be readily obtained from the master equation (8). Clearly, the diagonal element $\rho_{l,l}(t)$ gives the probability to find the photon in the $l$-th cavity of the lattice at time $t$, whereas $\langle vac | \hat{\rho} | vac \rangle$ corresponds to the probability that the photon has been destructed, with ${\rm Tr} (\hat{\rho})=\sum_l \rho_{l,l}+\rho_{vac,vac}=1$.
 In particular, it turns out that the dynamics of the density matrix elements $\rho_{n,m}$ is decoupled from the other elements and described by the set of coupled equations
\begin{eqnarray}
\frac{d}{dt} \rho_{n,m} & = & -\frac{\Gamma}{2} \left( \rho_{n,m+1}+\rho_{n,m-1}+\rho_{n+1,m}+\rho_{n-1,m} \right) \\
&- & \left(2 \Gamma+\frac{\gamma_n+\gamma_m}{2} \right) \rho_{n,m} \equiv \mathcal{L}^{(1)} \rho_{n,m}\nonumber
\end{eqnarray}
where  $\mathcal{L}^{(1)}$ describes the Liouvillian superoperator in the single-particle subspace. 
Likewise, the evolution equations of the density matrix elements $\rho_{n,vac}$ are decoupled from other density matrix elements and read
\begin{eqnarray}
\frac{d}{dt} \rho_{n,vac} & = &  -\frac{\Gamma}{2}( \rho_{n+1,vac}+\rho_{n-1,vac})-\left( \frac{\gamma_n}{2}+\Gamma \right) \rho_{n,vac}  \nonumber \\
& \equiv  & \sum_l \mathcal{H}_{n,l} \rho_{l,vac}
\end{eqnarray}
where 
\begin{equation}
\mathcal{H}_{n,l}=- \frac{\Gamma}{2} (\delta_{n,l+1}+\delta_{n,l-1})- \left( \frac{\gamma_n}{2}+\Gamma \right)  \delta_{n,l}
\end{equation}
are the matrix elements of the Wick-rotated non-Hermitian effective Hamiltonian, i.e. of the operator $\hat{\mathcal{H}}=-i \hat{H}_{eff}$, in the single-particle sector of Fock space. Note that, since $\hat{H}_{eff}$ is anti-Hermitian, the Wick-rotated operator $\hat{\mathcal{H}}$ is Hermitian and, remarkably, it corresponds to the Hermitian Hamiltonian of the ordinary one-dimensional Anderson model with uncorrelated diagonal disorder.

Finally, the evolution equation for the density matrix element $\rho_{vac,vac}$ reads
\begin{equation}
\frac{d}{dt} \rho_{vac,vac}= \Gamma \sum_n \left( 2 \rho_{n,n}+\rho_{n,n+1}+\rho_{n+1,n} \right)+ \sum_n \gamma_n \rho_{n,n}.
\end{equation}
Note that from Eqs.(11) and (14) it readily follows
\begin{equation}
\rho_{vac,vac}(t)=1-\sum_n \rho_{n,n}(t)
\end{equation}
as it should be owing to the property ${\rm Tr}(\hat{\rho})=1$. The dynamics of the density matrix elements is in some sense rather trivial: regardless of the initial excitation of the system, either in a coherent or mixture superposition of states, the asymptotic attractor of the dynamics  is the stationary state $\hat{\rho}_{\infty}$ of the Liouvillian superoperator, $\mathcal{L} \hat{\rho}_{\infty}=0$, which is trivially the vacuum state $\hat{\rho}_{\infty}=| vac \rangle \langle vac |$, i.e. 
\begin{equation}
(\rho_{\infty})_{n,m}=0 \; , \; (\rho_{\infty})_{n,vac}=0 \; , \; (\rho_{\infty})_{vac,vac}=1.
\end{equation}
However, the transient decay dynamics of the matrix elements $\rho_{n,m}(t)$ and $\rho_{n, vac}(t)$ toward the stationary state is nontrivial and displays the phenomenon of spectral localization without dynamic localization, previously observed in the Anderson model 2) with mixed dynamics (both Hamiltonian and dissipative) within an effective non-Hermitian approach \cite{r47,r48,r49}. In fact,  let us calculate the decay behavior of the density matrix elements $\rho_{n,vac}(t)$ and $\rho_{n,m}(t)$ by solving Eqs.(11) and (12). To this aim,
let us indicate by $u_l^{(\alpha)}$ and $\lambda_{\alpha}$ the eigenvectors and eigenvalues of the Hermitian Anderson Hamiltonian $\mathcal{H}$ [Eq.(13)], i.e. $\mathcal{H}u_n^{(\alpha)} = \lambda_{\alpha} u_n^{(\alpha)}$,
 where $\alpha$ is the eigenstate index. The general results of Anderson localization in the Hermitian case ensure that all the eigenstates $u_l^{(\alpha)}$ of $\mathcal{H}$ are exponentially localized, i.e. they decay exponentially in space as $l \rightarrow \pm \infty$, and the corresponding eigenenergies $\lambda_{\alpha}$ strictly satisfy the inequality
\begin{equation}
\lambda_{\alpha} < 0
\end{equation}
since $\gamma_n, \Gamma >0$. The most general solution to Eq.(14) is given by a superposition of the exponentially-localized eigenstates $u_n^{(\alpha)}$, whose amplitudes decay in time with a lifetime $1 / \lambda_{\alpha}$, i.e.
\begin{equation}
\rho_{n,vac}(t)=	\sum_{\alpha} C_{\alpha} u_n^{(\alpha)} \exp( \lambda_\alpha t).
\end{equation}
The complex constants $C_{\alpha}$ entering in Eq.(18) are determined by the initial values $\rho_{n,vac}(0)$, namely
\[
C_{\alpha}=\frac{\sum_n u_n^{(\alpha)} \rho_{n,vac}(0)} {\sum_n (u_n^{(\alpha)})^2}.
\]
 To calculate the solution to Eq.(11), let us observe that the eigenstates and corresponding eigenvalues of the Liouvillian superoperator $\mathcal{L}^{(1)}$ can be readily obtained from those of $\mathcal{H}$ by separation of variables, namely one has
\begin{equation}
\mathcal{L}^{(1)} u_n^{(\alpha)} u_m^{(\beta)}=(\lambda_{\alpha}+\lambda_{\beta}) u_n^{(\alpha)} u_m^{(\beta)}.
\end{equation}
 This yields
\begin{equation}
\rho_{n,m}(t)=	\sum_{\alpha, \beta} C_{\alpha, \beta} u_n^{(\alpha)} u_m^{(\beta)} \exp( \lambda_\alpha t+\lambda_{\beta} t)
\end{equation}
with some complex constants $C_{\alpha, \beta}$, which are determined by the initial values $\rho_{n,m}(0)$, namely
\[
C_{\alpha, \beta}=\frac{\sum_{n,m} u_n^{(\alpha)}  u_m^{(\beta)} \rho_{n,m}(0)} {\sum_{n,m} (u_n^{(\alpha)})^2 (u_m^{(\beta})^2}.
\]
 Clearly, as anticipated the matrix elements $\rho_{n,m}(t)$ and $\rho_{n,vac}(t)$ vanish as $t \rightarrow 0$, while simultaneously $\rho_{vac,vac}(t) \rightarrow 1$. {\color{black} {However, according to Eqs.(18) and (20) the decay dynamics is not monotonous  and is characterized by a sequence of "jumps" at around some times $t_1$ $t_2$, ..., $t_k$,.. where the excitation hops from one  exponentially-localized eigenstate with a shorter lifetime to another one with a longer  lifetime, located at some distance $d$ which can exceed the typical localization length of the eigenstates. 
 The jump dynamics and the resulting spatial spreading pattern in 
 the lattice, leading to dynamical delocalization, is very similar to the one discussed in details in previous works \cite{r47,r48,r49}. Between one jump and the next one, i.e. in the time interval ($t_{k+1}-t_k)$, the decay dynamics is approximately exponential with the rate of the instantaneously dominant eigenmode. Typically, the time interval $\Delta t_k$ required for the excitation to hope from one localized eigenstate to the other one at around the time instant $t_k$ is much shorter than the time interval $(t_{k+1}-t_{k})$ between two successive jumps. Also, the
 time interval $(t_{k+1}-t_k)$ rapidly increases as $k$ increases, i.e. the jumps become less and less frequent as time increases. Such a behavior basically stems from the fact that the difference between the lifetimes of the occupied eigenstates usually diminishes as time increases, so that the next dominant eigenstates take longer and longer times to replace the previous dominant eigenstates. An example of such a dynamical behavior will be illustrated in Fig.2 discussed below.}} The jump dynamics is determined by the
magnitude of the projection coefficients $C_{\alpha}$, $C_{\alpha,\beta}$ of the initial state on the eigenvectors of $\mathcal{H}$, and by the decay rate of each eigenstate \cite{r49}. {\color{black} The delocalization phenomenon arising from the jumps is observable for rather arbitrary (but not all) initial excitations of the system. For example, if the system is initially prepared in a pure state and exactly in an eigenstate $u_n^{(\alpha)}$ (or in a finite superposition of eigenstates), so as only a finite number of projection coefficients $C_{\alpha, \beta}$ are non-vanishing, delocalization is clearly not observed. Conversely, for single-site excitation of the system with almost sure probability the projection coefficients do not exactly vanish for quite arbitrary values of $\alpha$ and $\beta$, even though they can become extremely small when the eigenstates are localized far apart from the initially excited site: delocalization via jump dynamics is expected in this case. 
 It should be nevertheless mentioned that, in presence of noise in the system, even though the initial preparation of the system excites a finite number of eigenstates of  $\mathcal{H}$, excitation of other eigenstates with long lifetimes can be triggered even by a small noise (perturbation), and thus noise can restore delocalization via mode jumps.} 
To sum up, while the dissipative dynamics drives the system toward the vacuum state, both populations $\rho_{n,n}(t)$ and coherences $\rho_{n,m}(t)$, $\rho_{n,vac}(t)$ delocalize in the lattice via a sequence of jumps, despite all eigenstates of the Liouvillian superoperator are exponentially localized in space: in other words, we have coexistence of dynamical delocalization and spectral localization. We stress that such a dynamical delocalization process is very different than the delocalization observed in the dephasing Anderson model, where dynamical delocalization and the decay toward the maximally-mixed stationary state corresponds to spectral delocalization (see Appendix A).\\
 \begin{figure*}
  \includegraphics[width=\linewidth]{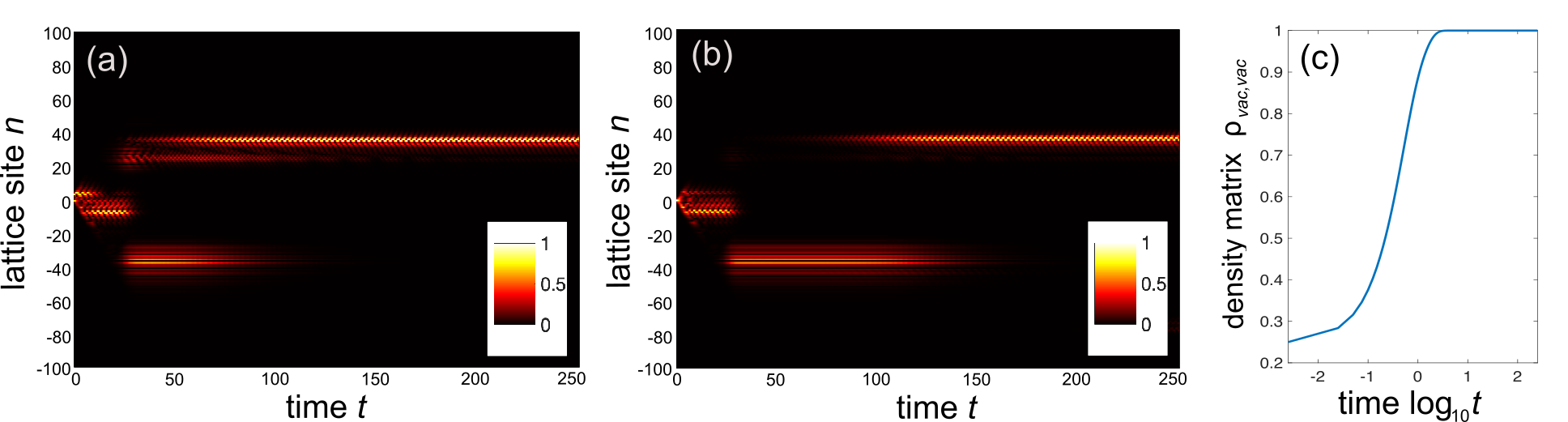}
  \caption{Dynamical evolution in the purely-dissipative Anderson model for a dissipative coupling $\Gamma=2$ and uncorrelated stochastic losses $\gamma_n$, taken from a uniform distribution in the range $(0,6)$. The system is initially prepared in the mixed state defined by Eq.(21) in the main text. Panel (a) shows, on a pseudocolor map, the temporal evolution of the normalized diagonal matrix elements $\tilde{\rho}_{n,n}$ (photon number) at various lattice sites, whereas panel (b) depicts the temporal evolution of the modulus of normalized coherences $\tilde{\rho}_{n,vac}$. Note the appearance of jumps between localized states located far apart one another. Panel (c) shows the evolution of the density matrix element $\rho_{vac,vac}(t)$, clearly indicating the fast convergence of the density operator toward the stationary state $\hat{\rho}_{\infty}=| vac \rangle \langle vac|$, corresponding to the vacuum state.}
\end{figure*}
As an illustrative example, let us consider the evolution dynamics of the density matrix elements for the initial state $\hat{\rho}(0)$ corresponding to the mixture of the two pure states
\begin{equation}
| \psi_1 \rangle= \frac{1}{ \sqrt{2}} \left( | vac \rangle + |-1 \rangle \right) \; ,\;\; | \psi_2 \rangle= | 3 \rangle
 \end{equation}
 with equal probabilities $p_{1}=p_{2}=1/2$. The pure state $|\psi_1 \rangle$ corresponds to the coherent superposition of states describing one photon at site $l=-1$ and no photon in the lattice, whereas the state $|\psi_2 \rangle$ corresponds to one photon localized at site $l=3$ in the lattice. The nonvanishing elements of the density matrix at initial time $t=0$ are thus $\rho_{
vac,vac}=1/4$, $\rho_{vac,-1}=\rho_{-1,vac}=1/4$, $\rho_{-1,-1}=1/4$ and $\rho_{3,3}=1/2$. 
 The time evolution of the density matrix has been obtained by numerical integration of Eqs.(11) and (12) using an accurate variable-step fourth-order Runge-Kutta method assuming a dissipative coupling rate $\Gamma=2$ and uncorrelated stochastic cavity losses $\gamma_l$ taken from a uniform distribution in the interval $(0,6)$. The lattice size has been assumed wide enough ($L=100$) so as to avoid edge effects over the largest propagation time $t=250$. The main results are depicted in Fig.2. Since the density matrix elements, with the exception of $\rho_{vac,vac}(t)$, rapidly decay toward zero, to highlight the jump dynamics in the figure the temporal behavior of normalized photon number $\tilde{\rho}_{n,n}(t) = \rho_{n,n}(t) / \sum_n \rho_{n,n}(t)$ and coherences $\tilde{\rho}_{n,vac}(t)= \rho_{n,vac}(t)/ \sqrt{\sum_n | \rho_{n,vac}(t)|^2}$ are plotted [Figs.2(a) and (b)], along with the temporal evolution of $\rho_{vac,vac}(t)$ [Fig.2(c)]. The plots clearly show that, while the density matrix rapidly converges toward the stationary state, corresponding to the vacuum state, delocalization of both population (photon number) and coherences is observed in the form of quantized jumps between localized eigenstates of the Anderson Hamiltonian $\mathcal{H}$.

\section{Localization in the purely-dissipative Anderson model: general analysis}
The coexistence of spectral localization of the Liouvillian superoperator $\mathcal{L}^{(1)}$ and dynamical delocalization for the density matrix elements in the single-particle subspace, shown in the previous section, is a general property that is valid beyond the single-particle limit. In fact, for open systems described by an Hamiltonian  {\it quadratic} in bosonic operators and with jump operators {\em linear} in bosonic operators --like in the purely-dissipative Anderson model-- the Lindblad master equation is exactly solvable and the Liouvillian $\mathcal{L}$ can be cast in a normal form \cite{r53}. To highlight the localization features of the model in the general case, for our purposes it is convenient to see the dynamics of the system monitoring the evolution of the mean values of first and second moments of the operators, i.e. $\alpha_n (t) \equiv \langle \hat{a}_n \rangle$  and correlations $\beta_{n,m} (t) \equiv \langle \hat{a}_n^{\dag} \hat{a}_m \rangle$, which are expressed by a set of linear differential equations (see for example \cite{r53,r54,r55}); technical details are given in the Appendix B. The evolution equations read
\begin{equation}
\frac{d \alpha_n}{dt}  =  - \frac{\Gamma}{2} \left( \alpha_{n+1}+\alpha_{n-1} \right)-\left( \Gamma+ \frac{\gamma_n}{2} \right) \alpha_n = \sum_l \mathcal{H}_{n,l} \alpha_l
\end{equation}
and
\begin{eqnarray}
\frac{d \beta_{n,m}}{dt} & = &  -\left( 2 \Gamma + \frac{\gamma_n+\gamma_m}{2} \right) \beta_{n,m} \nonumber \\
& - & \frac{\Gamma}{2} (\beta_{n+1,m}+\beta_{n-1,m}\beta_{n,m-1}+\beta_{n,m+1}) \\
& = & \mathcal{L}^{(1)} \beta_{n,m}. \nonumber
\end{eqnarray}
Interestingly, the evolution of the mean $\alpha_n$ is governed by the Anderson Hamiltonian $\mathcal{H}$, with the dynamics Wik-rotated in time like for single-particle case discussed in the previous section. Likewise, the evolution dynamics of the correlations $\beta_{n,m}(t)$ is governed by the Liouvillian $\mathcal{L}^{(1)}$ found in the single-particle sector discussed in the previous section. Therefore, following the the same line of discussion presented in the previous section, we can conclude that the purely-dissipative Anderson model displays dynamical delocalization for a rather arbitrary initial condition of the system, despite both $\mathcal{H}$ and $\mathcal{L}^{(1)}$ display exponential spectral localization. For example, let us assume that at initial time we have a mean of $N$ photons in the lattice site $n=0$ with some arbitrary photon statistics, while the other cavities are in the vacuum state, i.e. $\beta_{n,m}(0)=N \delta_{n,m}$. Then the dynamics of the normalized second-order moment
\begin{equation}
M_2(t)= \frac{\sum_{n} n^2 \beta_{n.n}(t)}{\sum_n \beta_{n,n}(t)},
\end{equation}
measuring the spreading of photons along the lattice,
 \begin{figure}
  \includegraphics[width=\linewidth]{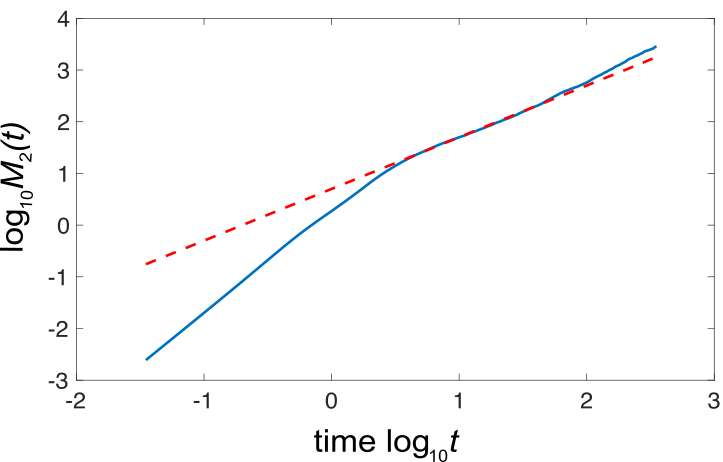}
  \caption{Numerically-computed behavior of $\overline{M_2(t)}$ versus time $t$, on a double log scale, in the purely-dissipative Anderson model for the same parameter values as in Fig.2. The behavior of the spreading law $\overline{M_2(t)}$ is obtained from Eq.(24) after averaging over 500 different realizations of disorder in the decay rates $\gamma_n$. The red dashed line corresponds to diffusive spreading with exponent $s=1$.}
\end{figure}
is unbounded as $t \rightarrow \infty$. Following the general analysis developed in Ref.\cite{r48}, the spreading dynamics at $t \rightarrow \infty$ can be well approximated  by the law $\overline{M_2(t)} \sim t^s$, where the overline denotes statistical average over the distribution of cavity losses $\gamma_n$. The value of the spreading coefficient $s= d \log (\overline{M_2(t))} /d \log t$ depends on the probability distribution density of the random variable $\gamma_n$; in particular, for a strong disorder with a uniform density distribution, one obtains diffusive spreading corresponding to $s=1$ \cite{r48}. This behavior is illustrated, as an example, in Fig.3. The figure depicts the numerically computed behavior of $\overline{M_2(t)}$ versus time $t$, on a double log scale, averaged over 500 different stochastic realizations of disorder with a uniform distribution and for the same parameter values as in Fig.2. As one can clearly see, after an initial transient the spreading of $\log \overline{M_2(t)}$ versus $\log t$ is well approximated by a line with slope $s$ close to one (dashed red curve in Fig.3). We remark that such a result holds regardless of the initial statistical distribution of photons in the system, i.e. it is valid for both classical and non-classical states of light, thus extending to the full quantum regime the stochastic jump dynamics of classical light, leading to diffusive-like transport, considered in previous studies within a semiclassical analysis \cite{r47,r48,r49}.

\section{Conclusions}
Anderson localization, i.e. the suppression of wave spreading in disordered lattices, is a rather universal mechanism of dynamical localization observed both in disordered quantum and classical systems. In conservative systems, this phenomenon arises from the delicate destructive interference  among multiply scattered waves, which localizes all wave functions (exponential spectral localization). However, when the system is open, i.e. when it exchanges energy and/or particles with the environment, the fate of Anderson localization can be deeply modified. For example, dephasing effects or measurements on the system restore delocalization \cite{r30,r31,r33,r40}, whereas disorder in dissipation can result in the coexistence of dynamical delocalization and spectral localization \cite{r47,r48,r49}. Previous Anderson models in open systems always involve mixed Hamiltonian (conservative) and Lindbladian (dissipative) dynamics, transport along the lattice being realized by conservative hopping. Here we have  introduced a purely-dissipative Anderson model, displaying an entirely Lindbladian dynamics \cite{r50}, and have shown that, while the system irreversibly decays toward the vacuum state -- the stationary state of the Liouvillian superoperator-- the transient dynamics displays diffusive-like behavior of both particle number and coherences along the lattice, in spite of the exponential spectral localization of the Liouvillian superoperator. The present results provide new insights into Anderson localization in dissipative systems and extend to the full quantum regime the intriguing possibility of coexistence of dynamical delocalization and spectral localization, recently predicted and observed in the framework of semiclassical models \cite{r47,r48,r49}. {\color{black} Finally, we mention that our analysis could be extended to investigate further intriguing dynamical effects that could arise in purely dissipative models with disorder. For example, one could consider the purely dissipative Anderson model in three dimensions on a cubic lattice, where the Hamiltonian displays mobility edges with extended eigenstates near the band center and localized eigenstates near the band edges. This means that some eigenfunctions $u_n^{(\alpha)}$ entering in Eq.(20) are extended, whereas some others are localized. Since the localized wave functions near one of the band edge have a longer lifetime (i.e.  a smaller decay rate $|\lambda_{\alpha}|$) than extended wave functions, one can envisage two competing types of delocalization: the "static" delocalization (spreading) phase at early times, originating from the extended nature of the Hamiltonian eigenstates near the band center, and the non-Hermitian-driven jump dynamical delocalization phase (like in the one-dimensional model) at longer times, originating from the hopping dynamics among localized eigenstates with longer lifetimes. Such an intriguing dynamical scenario could be observable also in one-dimensional models with disorder displaying mobility edges, such as in the extended Aubry-Andre models \cite{referee1,referee2,referee3}.}
\medskip

\medskip
\textbf{Acknowledgements} \par 
The author acknowledges the Spanish State Research Agency, through the Severo Ochoa
and Maria de Maeztu Program for Centers and Units of Excellence in R\&D (Grant No. MDM-2017-0711).

\medskip

%





  \appendix
  \renewcommand{\theequation}{A.\arabic{equation}}
\setcounter{equation}{0}
\small

\section{The dephasing Anderson model}
The dephasing Anderson model is obtained from Eqs.(6) and (7) by letting $\Gamma=\gamma_l=0$,and it is thus described by the effective non-Hermitian Hamiltonian
\begin{equation}
\hat{H}_{eff}=\hat{H}-\frac{i}{2} \gamma_{ph} \sum_l \hat{a}^{\dag}_{l} \hat{a}_l \hat{a}^{\dag}_{l} \hat{a}_l  \label{A1}
\end{equation}
where the Hermitian Hamiltonian $\hat{H}$ is given by Eq.(2). Contrary to the purely-dissipative Anderson model considered in Sec.2, the dephasing Anderson model conserves the number $N$ of particles (photons). 
To highlight the main features of this model, let us consider the dynamics in the single-particle sector of Fock space, so that the evolution of the system is described by the temporal evolution of the density matrix elements $\rho_{n,m}(t)=\langle n | \rho(t) | m \rangle$, where $|n\rangle=\hat{a}_n^{\dag} |vac \rangle$ is the single-particle Fock state that localizes one photon at the lattice site $n$. Using the commutation relations of bosonic operators, $ [\hat{a}_n,\hat{a}^{\dag}_m]= \delta_{n,m}$ and $ [ \hat{a}_n,\hat{a}_m]=[ \hat{a}_n^{\dag},\hat{a}_m^{\dag}]=0$, one readily obtains the following evolution equations
\begin{eqnarray}
 \frac{d}{dt} \rho_{n,m} & = & -iJ(\rho_{n,m+1}+\rho_{n,m-1}-\rho_{n+1,m}-\rho_{n-1,m}) \nonumber \\
 & + & i ( \delta \omega_m-\delta \omega_n) \rho_{n,m}-\gamma_{ph} \rho_{n,m} (1-\delta_{n,m})  \label{A2} \\
&  \equiv  &\mathcal{L}^{(1)} \rho_{n,m}  \nonumber
\end{eqnarray}
where $\delta_{n,m}$ is the Kronecker delta function. It can be readily shown that the steady-state solution $\hat{\rho}_{\infty}$ to Eq.(\ref{A2}), corresponding to an {\em extended} eigenvector of the one-particle Liouvillian superoperator $\mathcal{L}^{(1)}$ with zero eigenvalue, is the maximally-mixed state 
\begin{equation}
\hat{\rho}_{\infty}=\frac{1}{2L+1} \sum_l  | l \rangle \langle l | ,
\label{A3}
\end{equation}
i.e.$(\rho_{\infty})_{n,m}=\delta_{n,m}/(2L+1)$,
whatever is the Hermitian hopping amplitude $J$ and the cavity resonance detunings $\delta \omega_n$. This corresponds to a full and uniform delocalization of the photon along the lattice in a maximally mixed state, indicating that the pure dephasing terms fully destroy Anderson localization and drives the system toward the maximally-mixed state. The dynamical delocalization is here clearly associated to the existence of the extended state, i.e. breakdown of spectral localization, of the Liouvillian superoperator $\mathcal{L}^{(1)}$.\\ 
Finally, we mention that the semiclassical limit of the dephasing Anderson model, obtained by neglecting the quantum jump terms in the master equation and thus described by the evolution dynamics of the effective non-Hermitian Hamiltonian (A1) solely, does not destroy Anderson localization since the dephasing terms in the non-Hermitian hamiltonian just introduce a spatially-uniform decay rate of excitation at the various lattice sites. {\color{black} This is in contrast with the pure-dissipative Anderson model discussed in the main text, where the main features of Anderson localization are the same in the Lindbladian and effective non-Hermitian descriptions.}

  \renewcommand{\theequation}{B.\arabic{equation}}
\setcounter{equation}{0}
\small

\section{Evolution equations for the first and second moments}
In this Appendix we derive the evolution equations for the first and second moments of bosonic operators, $\alpha_n(t)= \langle \hat{a}_n \rangle$ and $\beta_{n,m}(t)= \langle \hat{a}_n^{\dag} \hat{a}_m \rangle$, given by Eqs.(22) and (23) in the main text. To this aim, let us observe that, since the system is purely dissipative, i.e. $\hat{H}=0$, for any time-independent operator $\hat{A}$ one has
\begin{equation}
\frac{d}{dt} \langle \hat{A} \rangle =  {\rm Tr} \left(  \frac{d \hat{\rho}}{dt} \hat{A} \right)=\sum_l \Gamma \; {\rm Tr} \left(   \mathcal{D} [ \hat{z}_l ] \hat{\rho} \hat{A} \right) + \sum_l \gamma_l 
\; {\rm Tr} \left( 
\mathcal{D} [ \hat{a}_l  ]\hat{\rho} \hat{A} \right)
\end{equation}
where 
\begin{equation}
\hat{z}_l= \hat{a}_{l+1}+\hat{a}_l
\end{equation}
and
\begin{equation}
\mathcal{D} [ \hat{o} ] \rho = \hat{o} \rho \hat{o}^{\dag}-\frac{1}{2} \left( \hat{o}^{\dag} \hat{o} \hat{\rho}+ \hat{\rho} \hat{o}^{\dag} \hat{o} \right). \label{B6}
\end{equation}
Using Eq.(\ref{B6}), it can be readily shown that
\begin{equation}
{\rm Tr} \left( \mathcal{D} [ \hat{o}] \hat{\rho} \hat{A} \right)=\frac{1}{2} {\rm Tr}\left( \hat{\rho} [ \hat{o}^{\dag} \hat{o} , \hat{A} ] \right)+ {\rm Tr}\left( \hat{\rho}  \hat{o}^{\dag} [ \hat{A},\hat{o} ] \right)
\end{equation}
so that the determination of the evolution equation of the mean value $\langle \hat{A} \rangle$ entails to calculate the commutators $[ \hat{o}^{\dag} \hat{o} , \hat{A} ]$ and $ [ \hat{A},\hat{o} ]$ for the two jump operators $\hat{o}=\hat{z}_l=\hat{a}_l+\hat{a}_{l+1}$ and $\hat{o}=\hat{a}_l$.\\
Let us first assume $\hat{A}=\hat{a}_n$, and let us set $\alpha_n(t) \equiv \langle \hat{a}_n \rangle$.  Taking into account that
\begin{eqnarray}
 \; [ \hat{z}^{\dag}_l \hat{z}_l , \hat{a}_n ]   & = &    -(\hat{a}_{n}+\hat{a}_{n+1}) \delta_{n,l}-(\hat{a}_n+\hat{a}_{n-1}) \delta_{n,l+1} \\ 
\;  [ \hat{a}^{\dag}_l \hat{a}_l , \hat{a}_n ]  & = &   -\hat{a}_n \delta_{n,l} \\
\; [\hat{a}_n, \hat{z}_l] & = & [\hat{a}_n, \hat{a}_l]=0 
\end{eqnarray}
from Eqs.(B.1-B.7) one readily obtains 
\begin{equation}
\frac{d \alpha_n}{dt}=- \frac{\Gamma}{2} \left( \alpha_{n+1}+\alpha_{n-1} \right)-\left( \Gamma+ \frac{\gamma_n}{2} \right) \alpha_n
\end{equation}
which is Eq.(22) given in the main text.\\
To calculate the temporal evolution of the second moments, let us assume $\hat{A}= \hat{a}_n^{\dag} \hat{a}_m$ and let us set $\beta_{n,m} \equiv \langle \hat{a}_n^{\dag} \hat{a}_m \rangle$.
 Taking into account that
\begin{eqnarray}
 \; [ \hat{z}^{\dag}_l \hat{z}_l , \hat{a}^{\dag}_n \hat{a}_m]   & = & (\delta_{l,n}-\delta_{l,m-1}+\delta_{l,n-1}-\delta_{l,m}) \hat{a}^{\dag}_{n} \hat{a}_m  \\
 &+ &  \delta_{l,n} \hat{a}^{\dag}_{n+1} \hat{a}_m 
 +  \delta_{l,n-1} \hat{a}^{\dag}_{n-1} \hat{a}_m \nonumber \\
 & - &  \delta_{l,m} \hat{a}^{\dag}_{n} \hat{a}_{m+1}- \delta_{l,m-1} \hat{a}^{\dag}_{n} \hat{a}_{m-1}  \nonumber   \\ 
\;  [ \hat{a}^{\dag}_l \hat{a}_l , \hat{a}_n^{\dag} \hat{a}_m ]  & = &   (\delta_{l,n}-\delta_{l,m} ) \hat{a}^{\dag}_n \hat{a}_m  \\
\; [ \hat{a}_n^{\dag} \hat{a}_m , \hat{z}_l] & = & -(\delta_{l,n}+\delta_{l,n-1}) \hat{a}_m \\
\;  [ \hat{a}_n^{\dag} \hat{a}_m , \hat{a}_l]& = & - \delta_{l,n} \hat{a}_m
\end{eqnarray}
from Eqs.(B.1-B.4) and (B.9-B.12) one obtains
\begin{eqnarray}
\frac{d \beta_{n,m}}{dt} & = &  -\left( 2 \Gamma + \frac{\gamma_n+\gamma_m}{2} \right) \beta_{n,m} \nonumber \\
& - & \frac{\Gamma}{2} (\beta_{n+1,m}+\beta_{n-1,m}\beta_{n,m-1}+\beta_{n,m+1})
\end{eqnarray}
which is Eq.(23) given in the main text.

\end{document}